\documentclass[amsmath,amssymb,showkeys,superscriptaddress,aps,prd,11pt,onecolumn]{revtex4-2}
\usepackage{graphicx}
\usepackage[utf8]{inputenc}
\usepackage[caption=false]{subfig}
\usepackage{multirow}
\usepackage{appendix}
\usepackage{graphicx,epstopdf}

\usepackage{colortbl}
\usepackage{xcolor}
\usepackage[colorlinks=true, urlcolor=blue, linkcolor=blue, citecolor=blue]{hyperref}
\usepackage{float}

\begin{document}

\title{Reappraisal of rho meson in nuclear matter by inverse QCD sum rules method}
\author{Halil Mutuk}%
\email[]{hmutuk@omu.edu.tr}
\affiliation{Department of Physics, Faculty of Sciences, Ondokuz Mayis University, 55200, Samsun, Türkiye}

\begin{abstract}
We present a comprehensive reappraisal of the in-medium properties of the rho meson using the inverse QCD sum rules (QCDSR) formalism, offering a novel, model-independent approach to studying hadronic modifications in nuclear matter. Unlike conventional QCDSR, which rely on a predefined pole+continuum structure, the inverse method reconstructs the spectral function directly from the operator product expansion (OPE), eliminating assumptions about the spectral ansatz. To the best of our knowledge, this is the first application of the inverse QCDSR method to the rho meson in nuclear matter. Our analysis reveals a significant reduction in the rho meson mass, consistent with previous theoretical predictions, and highlights the crucial role of medium-induced modifications, including condensate suppression and factorization-breaking effects. Furthermore, we assess the sensitivity of our results to the factorization assumption and higher-dimensional condensates, demonstrating the necessity of refining nonperturbative contributions for an accurate description of in-medium hadron properties. Our findings establish the inverse QCDSR method as a robust alternative to conventional spectral analysis techniques, providing a systematically controlled framework for exploring strongly interacting matter under extreme conditions. These results offer important theoretical benchmarks for lattice QCD simulations and heavy-ion collision experiments, shedding light on the restoration of chiral symmetry and the evolution of hadronic matter in dense environments.
\end{abstract}

\maketitle

\section{Rho Meson}\label{introduction}

A primary objective of contemporary nuclear and high-energy physics is to investigate the behavior of nuclear matter under extreme conditions. This topic is crucial for high-energy experiments, where experimental observables are influenced by medium modifications. At low temperatures and densities, quarks and gluons—the fundamental building blocks of strongly interacting matter—form hadrons due to the confinement mechanism. Additionally, the observation of very light mesons, such as pions and kaons, suggests the presence of chiral symmetry, which exhibits spontaneous breaking. This symmetry is approximately realized in the QCD Lagrangian. Another key indication of spontaneously broken chiral symmetry in the vacuum is the absence of chiral partners with equal masses.

The study of vector mesons, such as the rho meson, provides important insights into the fundamental properties of nuclear matter under various conditions. The rho meson plays a critical role in exploring the behavior of the strong nuclear force and its implications in dense and hot nuclear environments. Vector mesons are sensitive probes of medium modifications, which include changes in their mass, width, and spectral functions when embedded in nuclear matter. Such modifications are central to understanding phenomena in QCD and the interplay between hadronic excitations and medium interactions. Moreover, the rho meson is regarded as a promising candidate for observing potential indications of chiral symmetry restoration. This assertion is based on the fact that the rho meson shares quantum numbers with the photon. As a result, the rho meson can decay into a dilepton pair via a virtual photon. If this decay occurs within a hot and dense medium, the resulting dileptons carry information about the in-medium properties of the rho meson.

The rho meson exhibits several key properties when situated in nuclear matter, including changes in mass, width, and spectral density. These properties are influenced by the nuclear environment and interactions with other particles. For instance, it was shown in Ref. \cite{Brown:1991kk} that the vector meson mass should decrease in nuclear matter according to a scaling law, which may be a result of chiral symmetry restoration. The rho meson mass decreases in nuclear matter, with studies indicating a reduction of about 20\% at nuclear matter density \cite{Shakin:1993ta,Asakawa:1993pq}. This mass reduction is generally observed to increase with nuclear density and is supported by various models and approaches, including the vector dominance model and QCDSR formalism \cite{Asakawa:1992ht,Asakawa:1993pq,Ko:1994hc,Kondratyuk:1998ec,Teodorescu:2001ja}. The mass shift is attributed to changes in the quark condensate and interactions with the nuclear medium.

The width of the rho meson is also affected, typically narrowing by approximately 35\% at nuclear matter density \cite{Shakin:1993ta,Asakawa:1993pq}. This narrowing is linked to modifications in the meson interactions with the surrounding nuclear environment. Additionally, the rho meson experiences substantial broadening of its spectral density in dense nuclear matter, influenced by baryon resonances such as $N^\ast(1520)$ and $N^\ast(1720)$ \cite{Teodorescu:2001ja}. This broadening exhibits nontrivial momentum dependence. The rho meson peak in the spectral function shifts to smaller invariant masses as density increases, accompanied by a reduction in peak strength and the emergence of a low-mass peak that also shifts downward with increasing density \cite{Asakawa:1992ht,Asakawa:1993pq,Asakawa:1993pm,Ko:1994hc}. Further studies have indicated that the rho meson in nuclear matter may undergo complex modifications that cannot be fully described by a simple mass shift and broadening \cite{Rapp:1997fs,Klingl:1997kf,Peters:1997va,Post:2000qi,Post:2003hu}. These studies revealed an enhancement of the spectral strength in the low-energy region below the original rho meson peak.

The QCDSR technique \cite{Shifman:1978bx,Shifman:1978by} is a powerful theoretical tool for studying the non-perturbative aspects of QCD, particularly for understanding the properties of hadrons. Based on the QCD Lagrangian, QCDSR is one of the most frequently used methods for investigating hadron properties. The QCDSR formalism has been widely applied to analyze the mass of the rho meson at finite nuclear density \cite{Hatsuda:1991ez,Asakawa:1992ht,Asakawa:1993pq,Hatsuda:1995dy,Jin:1995qg,Leupold:1997dg,Klingl:1997kf,Kwon:2008vq,Mishra:2014rha,Kumar:2018pqs}, consistently supporting a universal decrease in mass. The results of the QCDSR technique are in agreement with a negative mass shift of the rho meson in nuclear matter.

As mentioned above, the rho meson undergoes significant modifications in nuclear matter, including a reduction in mass and width, broadening of spectral density, and shifts in its spectral function. These changes are primarily due to interactions with the nuclear medium and are influenced by factors such as quark condensate reduction and coupling with other particles. Understanding these properties is essential for exploring the behavior of the rho meson in dense environments. In this work, we reevaluate the properties of the rho meson in nuclear matter using the inverse QCDSR formalism. By solving for the spectral function directly, we eliminate ambiguities associated with continuum threshold choices and explore the impact of medium modifications with unprecedented precision.

This work is arranged as follows: in Section \ref{formalism}, we briefly explained conventional QCDSR method and introduced the inverse QCDSR method in detail. Section \ref{numerical} is devoted to the numerical analysis of the spectroscopic parameters of rho meson in nuclear medium. Section \ref{final} is reserved for concluding remarks.

\section{Inverse Problem of QCD sum rules formalism} \label{formalism}
\subsection{Conventional QCDSR Approach}

As a non-perturbative method, QCDSR establish a connection between hadronic properties, expressed through a set of parameters, and the non-perturbative structure of QCD, represented by vacuum condensates. This method begins by initializing the system in the asymptotic freedom state of quarks, a configuration valid at short distances. The system is then gradually extended to long distances, where bound states in QCD emerge. The need for a non-perturbative approach arises because the asymptotic freedom state degrades during this process, leading to resonances that correspond to bound quark states confined within hadrons. The breakdown of asymptotic freedom gives rise to non-perturbative phenomena within the QCD vacuum, manifesting as non-zero values of quark and gluon density operators in the vacuum.

The QCDSR approach relates certain low-energy quantities, which are not directly accessible through QCD calculations, to high-energy expressions. These high-energy expressions can be computed using OPE in terms of quark and gluon degrees of freedom. Non-perturbative effects are embedded in the form of various quark and gluon condensates.

A QCDSR calculation starts with the correlation function
\begin{equation}
\Pi=i \int d^4 x e^{i q x} \langle 0 \vert {\cal T}  \left[ j(x) j^\dagger(0) \right] \vert 0 \rangle, \label{eq:corfunc}
\end{equation}
where $j(x)$ is an operator (interpolating current) consisting of quark and gluon fields, which when applied to the vacuum can create the hadron of interest, and ${\cal T}$ is a time-ordering operator. For the case $q^2 > 0$, if the expression of the unit operator written as a sum over hadron states is inserted between the operators in the correlation function Eq. \ref{eq:corfunc}, the following expression can be obtained:
\begin{eqnarray}
\Pi&=&\sum_h \langle 0 \vert j \vert h(q) \rangle \frac{1}{q^2-m_h^2} \langle h(q) \vert j^\dagger \vert 0 \rangle + \mbox{multiple hadron states}.  \label{eq:piphen} 
\end{eqnarray}
In the case of $-q^2 >\Lambda^2_{QCD}$, the majority of the contribution to the correlation function originates from the region $x\sim 0$. In this particular instance, the product of two operators can be expressed through OPE:
\begin{equation}
{\cal T} \left[ j(x) j^\dagger(0) \right] = \sum_d C_d(x) O_d,
\end{equation}
where $C_d(x)$ are the coefficients computable by perturbation theory, and $O_d$ are local operators with mass dimension $d$. Applying Fourier transformation, one can obtain:
\begin{equation}
\Pi = \sum_d C^f_d(q) \frac{\langle O_d \rangle}{q^d}, \label{eq:piqcd}
\end{equation}
where $\langle O_d \rangle$ are vacuum condensates that cannot be calculated with perturbation theory, except $(d=0)$, which includes contributions to the correlation function that can be computed by perturbation theory. To derive the sum rules, it is necessary to equate Eq. (\ref{eq:piphen}) and Eq. (\ref{eq:piqcd}). However, it should be noted that these expressions are derived for distinct values of $q^2$. To establish a connection between these two regions, the spectral representation of the correlation function can be employed:
\begin{equation}
\Pi(q^2) = \int_0^\infty \frac{\rho(s)}{s-q^2} + \mbox{polynomials in $q^2$}. \label{eq:spectral}
\end{equation}
The spectral density $\rho(s)$ can be obtained from Eq. (\ref{eq:piphen}). Once this is obtained, it can be inserted into Eq. (\ref{eq:spectral}), and an expression can be derived for the region $q^2<0$. Denoting the spectral density obtained from equation Eq. (\ref{eq:piphen}) as $\rho^{phen}(s)$ and the one obtained from Eq. (\ref{eq:piqcd}) as $\rho^{QCD}(s)$, one can write:
\begin{eqnarray}
\int_0^\infty ds \frac{\rho^{phen}(s)}{s-q^2} + \mbox{polynomials}&=& \int_0^\infty ds \frac{\rho^{QCD}(s)}{s-q^2} 
+ \mbox{polynomials}.
\end{eqnarray}
To get rid of the polynomials here, both sides can be sufficiently differentiated. However, in general, since the degree of the polynomials is unknown, one can perform the Borel transformation, which involves differentiating infinitely many times. The Borel transformation is defined as:
\begin{equation}\label{borel}
\Pi(M^2)=\mathcal{B}[\Pi(q^2)]\equiv 
\lim\limits_{\tiny \begin{matrix} q^2, n 
\rightarrow \infty \\ q^2/n = M^2 \end{matrix}} 
\frac{(q^2)^{n+1}}{(n)!}\Big(-\frac{d}{dq^2}\Big)^n\, \Pi(q^2)\,,
\end{equation}
where $M^2$ is the Borel mass parameter. The main effect of this transformation is to eliminate polynomials while simultaneously transforming:
\begin{equation*}
\frac{1}{s-q^2} \rightarrow e^{-\frac{s}{M^2}}.
\end{equation*}
After Borel transformation, the relationship between hadronic properties and QCD parameters can be expressed as follows:
\begin{eqnarray}
 \int_0^\infty \rho^{QCD}(s) e^{- \frac{s}{M^2}} &=& \sum_h \vert \langle 0 \vert j \vert h(q) \rangle \vert^2 e^{-\frac{m_h^2}{M^2}} + \mbox{multiple hadron states}.
\label{eq:sumrule1}
\end{eqnarray}
There are still infinitely many unknown parameters in this expression. Notice that, due to the exponential $e^{-\frac{m_h^2}{M^2}}$, the main contribution to the left-hand side will come from hadrons with mass at the ground state. In order to refine this approximation and parametrize the contributions of multiple hadron states, quark-hadron duality can be used. According to quark-hadron duality, for $s>s_0$, $\rho^{phen}(s) \simeq \rho^{QCD}(s)$, i.e., the spectral function of the continuum and higher states can be expressed as the spectral function of the OPE above a critical energy. For $s>s_{0}$, $\rho^{phen}(s)$ includes contributions from heavier states and multiple hadron states. Using quark-hadron duality, Eq. (\ref{eq:sumrule1}) can be written as:
\begin{equation}
\vert \langle 0 \vert j \vert h_t(q) \rangle \vert^2 e^{-\frac{m_{h_t}^2}{M^2}} = \int_0^{s_0} \rho^{QCD}(s) e^{-\frac{s}{M^2}}.
\label{eq:sumrule2}
\end{equation}
Here $h_t$ is the lowest mass hadron that can be created by the $j$ operator. Denoting the coupling (decay constant or pole residue) $\lambda$ of the current with the mass of the lowest mass of hadron as:
\begin{equation}
\lambda= \langle 0 \vert j \vert h_t(q) \rangle,
\end{equation}
Eq. (\ref{eq:sumrule2}) can be written as:
\begin{equation}
\lambda^2 e^{-\frac{m_{h_t}^2}{M^2}} = \int_0^{s_0} \rho^{QCD}(s) e^{-\frac{s}{M^2}},
\label{eq:sumrule3}
\end{equation}
which yields the sum rules of QCD. From these sum rules, hadronic properties, such as mass and decay constant, can be extracted. In the QCDSR approach, the use of some phenomenological inputs limits the accuracy of the method to about 10\% - 20\%. In addition to this, other sources of uncertainty such as factorization of higher-dimensional operators, truncation of the OPE at some order, working windows for continuum threshold and Borel parameter lay behind the formalism \cite{Leinweber:1995fn}.

The conventional approach to QCDSR analysis entails the identification of an appropriate `working window" which encompasses the continuum threshold $s_0$ and the Borel parameters, also referred to as auxiliary parameters. Given that the auxiliary parameters are not physical (they cannot be observed experimentally), the physical properties extracted from the sum rules must be either independent of these parameters or have a very weak dependence on them. To this end, it is essential to identify working windows in which physical quantities exhibit a negligible dependence on these auxiliary parameters. The continuum threshold $s_0$ is defined as the parameter associated with the first excited state of the ground state, which possesses the same quantum numbers as the hadron under consideration. In essence, the continuum threshold $s_0$ serves to distinguish the ground state contribution from continuum states and higher resonances in the correlation function. Estimates of the continuum threshold are contingent upon experimental data concerning the masses of the ground and first excited states of the hadronic state under investigation. For conventional hadrons, the parameters of the excited states can be obtained from experimental measurements or theoretical studies. In the context of exotic hadrons, there is considerable uncertainty about the availability of relevant experimental information. The Borel mass parameter interval is chosen to minimize the impact of higher resonances and the continuum, while also reducing the contribution of higher-order operators with negligible contributions. The Borel parameter $M^2$ should be selected as minimal as possible in order to obtain the contribution from the single resonance that lies at the lowest position. The exponential suppression factor given in Eq. (\ref{eq:sumrule3}) results from the Borel transformation, and the integral receives the main contribution at $s \simeq M^2$. At lower values of $M^2$, the integral gets the dominant contribution from lower $s$ values, and the contribution decreases at large values of $s$, $s \geq M^2$. Hence, the lowest single resonance dominates the integral, and quark-hadron duality is consistent. At large values of $M^2$, asymptotic freedom enters the game, and since the lowest lying resonance does not contribute significantly to the integral, quark-hadron duality is not reliable. Therefore, the Borel parameter $M^2$ should not be too large but not too small. A balance between the upper and lower values of $M^2$ is sought for reliable results, which means convergence of the series and dominance of the single pole contribution.

The other issue encountered with skepticism is parametrizing the spectral function with the ``pole + continuum" ansatz, where the continuum denotes the excited and scattered states, and the pole denotes the hadron in question. As mentioned in Ref. \cite{Gubler:2010cf}, while this assumption might work well when the low-energy part of the spectrum is dominated by a single pole and the continuum states only become important at higher energies, it is not certain if it will work well in other situations. In sum rule analysis, the pole contribution is defined as:
\begin{equation}
\text{PC}=\frac{\Pi(s_0, M^2)}{\Pi(\infty, M^2)}=
\left\{
	\begin{array}{ll}
		  \gtrsim 20,  & \mbox{for exotic hadrons}, \\
		 \gtrsim 50,  & \mbox{for conventional hadrons}.
	\end{array}
\right.
\end{equation}
The obtained sum rules are asked to satisfy the pole contribution. If there is not a suitable interval for the continuum threshold and Borel parameter that satisfies the pole contribution, the extracted observables may not be reliable, and one can conclude that there exists no sum rule of QCD for the corresponding state.

As mentioned above, QCDSR has achieved great success in studying quark masses, form factors, mass spectra, decay constants, decay widths, and magnetic moments of conventional and exotic hadrons. Notwithstanding, alternative formulations and approaches take place in the literature. The recent one is related to the inverse problem of QCDSR formalism with various applications \cite{Li:2020ejs,Li:2021gsx,Li:2022qul,Li:2022jxc,Xiong:2022uwj,Li:2023dqi,Li:2023yay,Li:2023ncg,Li:2023fim,Li:2024awx,Li:2024xnl,Li:2024fko,Mutuk:2024jvv}.

\subsection{Inverse QCDSR Approach}

In the inverse problem of QCDSR formalism, the OPE side of the correlation function is calculated in the standard way. The generic spectral density on the hadron side of the correlation function is treated as an unknown. The inverse QCDSR formalism aims to reconstruct the spectral density, which encodes information about hadronic resonances, directly from the OPE and the dispersion relation. The spectral density is treated as an unknown function to be reconstructed using the OPE input, solving for it as an inverse problem. No assumptions about a continuum threshold or specific resonance structure are required.

The unknown spectral density $\rho(y)$ is expanded using generalized Laguerre polynomials $L_n^{(\alpha)}(y)$:
\begin{equation}
\rho(y) = \sum_{n=1}^{N} a_n y^\alpha e^{-y} L_{n-1}^{(\alpha)}(y),
\end{equation}
where $N$ is the maximum polynomial degree, and $a_n$ are the expansion coefficients. The parameter $\alpha$ is chosen based on the boundary conditions of $\rho(y)$, such as its behavior at small $y$ and its asymptotic behavior, $y \to 0$.

The generalized Laguerre polynomials $L_n^{(\alpha)}(y)$ satisfy:
\begin{eqnarray}
\int_0^\infty y^\alpha e^{-y}L_m^{(\alpha)}(y)L_n^{(\alpha)}(y)dy=\frac{\Gamma(n+\alpha+1)}{n!}\delta_{mn},
\end{eqnarray}  
where $\Gamma$ is the gamma function. This property ensures that the polynomials form a complete and orthogonal basis over the interval $[0, \infty)$. Orthogonality simplifies the calculation of the expansion coefficients $a_n$, as it ensures minimal overlap between terms. This minimizes numerical errors and avoids redundancies in the representation. This orthogonality also ensures that each polynomial contributes independently to the representation of the spectral density $\rho(y)$.

The dispersion relation connecting the correlation function $\Pi(q^2)$ to the spectral density $\rho(y)$ is expressed as:
\begin{equation}
\int_{0}^\infty \frac{\rho(y)}{x - y} \, dy = \omega(x),
\end{equation}
where $x = p^2/\Lambda$ and $\omega(x)$ is determined from the OPE.  
By substituting the spectral density expansion into this equation, a matrix form emerges:
\begin{equation}
M \mathbf{a} = \mathbf{b},
\end{equation}
where $M$ contains integrals of Laguerre polynomials, $\mathbf{a}$ is the vector of coefficients $a_n$, and $\mathbf{b}$ comes from the OPE input. The coefficients are obtained using $\mathbf{a} = M^{-1} \mathbf{b}$ if $M$ is invertible. The existence of a solution ensures that the reconstructed $\rho(y)$ is consistent with theoretical inputs.

Generalized Laguerre polynomials are used to expand the unknown spectral density in the dispersion integral. Next, the coefficients of $1/(q^2)^m$ on both sides of the sum rule are equated to create a matrix equation. After that, the matrix equation can be solved to obtain an approximate solution for the spectral density. As mentioned before, this approach avoids traditional assumptions like continuum thresholds or Borel transformations.

\subsection{Rho meson in nuclear matter}
\begin{figure}[h!]
\centering
\includegraphics[width=10cm]{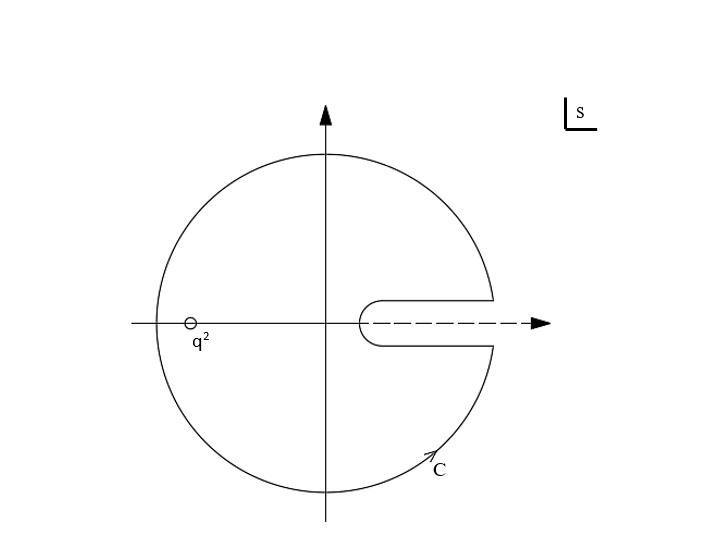}
\caption{Contour on the complex s plane. The dashed line shows potential poles and cuts of $\Pi(s)$.}
\label{contour}
\end{figure}

The rho meson has the quark current $J_\mu=(\bar u\gamma_\mu u-\bar d\gamma_\mu d)/\sqrt{2}$, which can be inserted into the two-point correlator:
\begin{eqnarray}
\Pi_{\mu\nu}(q^2)&=&i\int d^4xe^{iq\cdot x}
\langle 0|T[J_\mu(x)J_\nu(0)]|0\rangle =(q_\mu q_\nu-g_{\mu\nu}q^2)\Pi(q^2). \label{corfunc}
\end{eqnarray}
The vacuum polarization function $\Pi(q^2)$ follows the relation:
\begin{eqnarray}
\Pi(q^2)=\frac{1}{2\pi i}\oint \frac{\Pi(s)}{s-q^2} ds,\label{corfunc1}
\end{eqnarray}
where the contour shown in Fig. \ref{contour} is made up of a circle with a large radius $R$ and two horizontal lines cut along the positive real axis on the complex $s$ plane, above and below the branch.

The left-hand side of the $\Pi(q^2)$ function given in Eq. (\ref{corfunc1}) can be decomposed as:
\begin{equation}
\Pi^\text{OPE}(q^2)=\Pi^\text{Pert}(q^2) + \Pi^\text{Cond}(q^2) \label{corope}
\end{equation}
where for the rho meson in nuclear matter, the perturbative representation reads as \cite{Hatsuda:1992bv,Hatsuda:1995dy}:
\begin{eqnarray}
\Pi^\text{Pert}=-{1\over 8\pi^2} \left(1+{\alpha_s \over \pi}\right) 
    {\rm ln}\left({Q^2 \over \mu^2}\right) + {m_N \over 4 Q^4} A_2 \rho_N - {5 m_N^3 \over 12 Q^6} A_4 \rho_N, \label{mediumpert}
\end{eqnarray}
and $\Pi^\text{Cond}(q^2)$ reads as:
\begin{eqnarray}
\Pi^{\rm Cond}_{\rho_N} = \frac{1}{Q^4}m_q\langle\bar{q}q\rangle_{\rho_N} + \frac{1}{24Q^4}\left\langle\frac{\alpha_s}{\pi}G^2\right\rangle_{\rho_N} - \frac{112}{81Q^6}\pi\alpha_s\kappa\,\langle\bar{q}q\rangle_{\rho_N}^2. \label{mediumcond}
\end{eqnarray}
Here $m_N$ is the nucleon mass, $\rho_N$ denotes the nuclear density, $A_2$ and $A_4$ are factors which will be borrowed from quark-parton distributions, and $\kappa$ parametrizes deviations from the factorization assumption of higher-order vacuum condensates. In Eq. (\ref{mediumpert}), the regularization-scheme dependent term of ${\rm Im}\Pi^{\rm pert}(s)$ is dropped since it has no bearing on the quest for a resonance solution. The quark and gluon condensates in nuclear matter obey the relation \cite{Drukarev:1988kd}:
\begin{equation}\label{sina}
 \langle{\cal O}_i\rangle_{\rho_N}=\langle{\cal O}_i\rangle_0+\frac{\rho_N}{2 M_N}\langle{\cal O}_i\rangle_N+o(\rho_N),
\end{equation}
where condensates $\langle{\cal O}_i\rangle_N$ are related to the 
condensates $\langle{\cal O}_i\rangle_{\rho_N}$, $\rho_N$ is the density of nuclear matter, and $m_N$ is the nucleon mass.

The right-hand side of Eq. (\ref{corfunc1}) can be written as:
\begin{eqnarray}
\frac{1}{2\pi i}\oint ds\frac{\Pi(s)}{s-q^2}&=&
\frac{1}{\pi}\int_{0}^R ds\frac{{\rm Im}\Pi(s)}{s-q^2} + \frac{1}{2\pi i}\int_C ds\frac{\Pi^{\rm pert}(s)}{s-q^2},\label{corfunc2}
\end{eqnarray}
where the imaginary part of ${\rm Im}\Pi(s)$ will be regarded as unknown. This imaginary part of ${\rm Im}\Pi(s)$ includes nonperturbative dynamics from the lower $s$ region of the branch cut. The second term in Eq. (\ref{corfunc2}) is presented by replacing the numerator as $\Pi^{\rm pert}(s)$ since the perturbative calculation of $\Pi(s)$ is reliable for $s$ far away from physical poles. This replacement is in accordance with the OPE in Eq. (\ref{corope}).

Furthermore, the perturbative part $\Pi^\text{Pert}(q^2)$ in Eq. (\ref{corope}) can be written as an integral along the same contour such that:
\begin{eqnarray}
\Pi^{\rm OPE}(q^2) = \frac{1}{2\pi i}\oint ds\frac{\Pi^{\rm pert}(s)}{s-q^2} + \frac{1}{Q^4}m_q\langle\bar{q}q\rangle_{\rho_N} + \frac{1}{24Q^4}\left\langle\frac{\alpha_s}{\pi}G^2\right\rangle_{\rho_N} - \frac{112}{81Q^6}\pi\alpha_s\kappa\,\langle\bar{q}q\rangle_{\rho_N}^2.\label{corfunc3}
\end{eqnarray}

We get the sum rule by equating Eq. (\ref{corfunc2}) and Eq. (\ref{corfunc3}), 
\begin{eqnarray}
\frac{1}{\pi}\int_{0}^{R}ds\frac{{\rm Im}\Pi(s)}{s-q^2} &=& \frac{1}{\pi}\int_{0}^{R}ds\frac{{\rm Im}\Pi^{\rm pert}(s)}{s-q^2} + \frac{1}{(q^2)^2}m_q\langle\bar{q}q\rangle_{\rho_N} + \frac{1}{24(q^2)^2}\left\langle\frac{\alpha_s}{\pi}G^2\right\rangle_{\rho_N} \nonumber \\  &-& \frac{112}{81(q^2)^3}\pi\alpha_s\kappa\,\langle\bar{q}q\rangle_{\rho_N}^2,  \label{sumrule} 
\end{eqnarray}
where contributions of $\Pi^{\rm pert}(s)$ along the large circle $C$ and the renormalization scale $\mu$ have both canceled. In conventional QCDSR analysis, the spectral density is defined from the $\Pi(s)$ as $\rho(s) \equiv \frac{1}{\pi}{\rm Im}\Pi(s)$. A similar expression can be written by introducing the subtracted spectral density as:
\begin{equation}
\Delta\rho(s,\Lambda) = \rho(s) - \frac{1}{\pi}{\rm Im}\Pi^{\rm pert}(s)\left[1 - e^{-s/\Lambda}\right], \label{subtracted}
\end{equation}
where $\Lambda$ is the scale which characterizes the transition of ${\rm Im}\Pi(s)$ to the ${\rm Im}\Pi^{\rm pert}(s)$. The subtracted spectral density has some features with respect to changes in $s$. Looking at the term $1-e^{-s/\Lambda}$, one can see that it behaves like the leading order of $s$ and vanishes as $s$ approaches zero, $s \to 0$. The subtracted spectral density has distinct asymptotic behaviors:
\begin{itemize}
    \item For $s \ll \Lambda$: $\Delta\rho(s,\Lambda) \approx \rho(s) \sim s$ as $s\to 0$, capturing the non-perturbative resonance structure
    \item For $s \gg \Lambda$: $\Delta\rho(s,\Lambda) \to 0$ rapidly, as $\rho(s)$ approaches its perturbative form $\rho^{\text{pert}}(s)$
\end{itemize}
The transition scale $\Lambda$ governs the crossover between these regimes. If the sum rule expression given in Eq. (\ref{sumrule}) is written in terms of the subtracted spectral density, it is possible to extend the upper limit of the integral to infinity:
\begin{eqnarray}
\int_{0}^{\infty}\frac{\Delta\rho(s,\Lambda)}{s-q^2}ds &=& \int_{0}^{\infty}\frac{\Pi^{\rm pert}(s)e^{-s/\Lambda}}{s-q^2}ds + \frac{1}{(q^2)^2}m_q\langle\bar{q}q\rangle_{\rho_N} + \frac{1}{24(q^2)^2}\left\langle\frac{\alpha_s}{\pi}G^2\right\rangle_{\rho_N} \nonumber \\ &-& \frac{112}{81(q^2)^3}\pi\alpha_s\kappa\,\langle\bar{q}q\rangle_{\rho_N}^2. \label{sumrule2} 
\end{eqnarray}
It is worth mentioning that quark-hadron duality is not postulated for any finite value of $s$ while deriving this sum rule, contrary to the conventional QCDSR analysis—note that in standard QCDSR analysis, quark-hadron duality is assumed according to $s$.

The subtracted spectral density $\Delta \rho(s,\Lambda)$ is a dimensionless quantity and can be written in terms of $\Delta \rho(s/\Lambda)$. If the change of variables $x=q^2/\Lambda$ and $y=s/\Lambda$ is done, Eq. (\ref{sumrule2}) transforms as:
\begin{eqnarray}
\int_{0}^{\infty}\frac{\Delta\rho(y)}{x-y}dy &=& \int_{0}^{\infty}\frac{\Pi^{\rm pert}(s)e^{-y}}{x-y}dy + \frac{m_q\langle\bar{q}q\rangle_{\rho_N}}{\Lambda^2 x^2} + \frac{1}{24\Lambda^2 x^2}\left\langle\frac{\alpha_s}{\pi}G^2\right\rangle_{\rho_N} \nonumber \\ &-& \frac{112}{81\Lambda^3 x^3}\pi\alpha_s\kappa\,\langle\bar{q}q\rangle_{\rho_N}^2. \label{sumrule3} 
\end{eqnarray}

The boundary conditions:
\begin{eqnarray}
\Delta\rho(y) \sim y, ~\text{when} ~ y \to 0, \nonumber \\ 
\Delta\rho(y) \sim 0, ~\text{when} ~ y \to \infty, 
\end{eqnarray}
make it possible to expand generalized Laguerre polynomials. The stability analysis of the sum rule can be done where extracted parameters do not change heavily with respect to changes in $\Lambda$. It plays a similar role to the Borel mass parameter in conventional QCDSR analysis. More rigorous prescriptions can be found in Refs. \cite{Li:2020ejs,Li:2021gsx}.

\section{Results and Discussion}\label{numerical}

We use the following set of input for the analysis: $A_2=0.9$ and $A_4=0.12$ \cite{Hatsuda:1995dy}, $\langle \bar{q}q \rangle_{\rho_N}=\langle \bar{q}q \rangle + \frac{\sigma_N}{m_u+m_d}\rho_N$, $m_u+m_d=12~\text{MeV}$, $\sigma_N=(45 \pm 10) ~\text{MeV}$, $\langle \frac{\alpha_sGG}{\pi}\rangle_{\rho_N}=\langle \frac{\alpha_sGG}{\pi}\rangle-(0.65 \pm 0.15)~\text{GeV} \rho_N$, $\langle \bar{q}q \rangle=(-0.23)^3 ~\text{GeV}^3$, $\langle \frac{\alpha_sGG}{\pi}\rangle=(0.33)^4 ~\text{GeV}^4$, $m_q=6 ~\text{MeV}$, $\rho_N=(0.11)^3 ~\text{GeV}^3$ \cite{Furnstahl:1992pi,Jin:1992id,Jin:1993up,Cohen:1994wm}, $\alpha_s=0.5$, $1.0 \leqslant \kappa \leqslant 3.5$ \cite{Govaerts:1986ua,Leinweber:1995fn}.

In inverse QCDSR, $\Lambda$ is a characteristic scale that determines the transition from the non-perturbative resonance region to the perturbative QCD region. Unlike in conventional QCDSR (which uses a fixed continuum threshold $s_0$), the inverse method solves for the spectral density directly but still requires a cutoff scale $\Lambda$ to regulate the subtraction procedure. The key requirement is that $\Lambda$ should be chosen such that (i) the extracted hadronic properties (e.g., rho meson mass) do not strongly depend on $\Lambda$ and (ii) the spectral density remains positive and smooth, avoiding artificial oscillations.

\begin{figure}[h!]
\begin{center}
\includegraphics[totalheight=5cm,width=7cm]{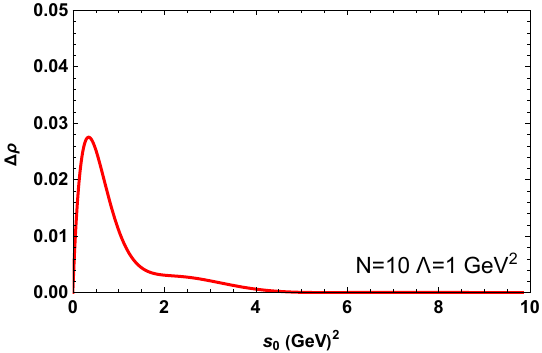}
\includegraphics[totalheight=5cm,width=7cm]{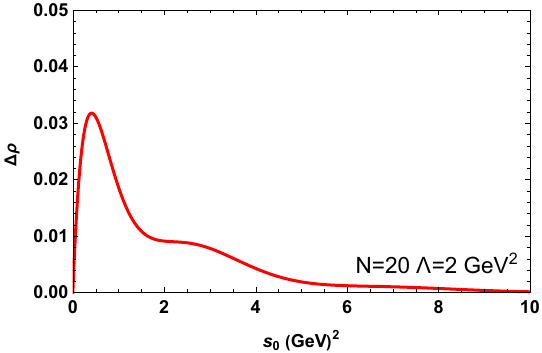}
\includegraphics[totalheight=5cm,width=7cm]{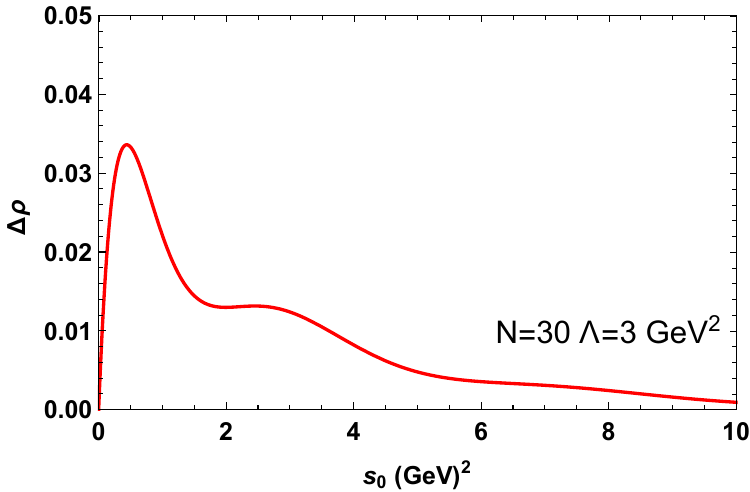}
\includegraphics[totalheight=5cm,width=7cm]{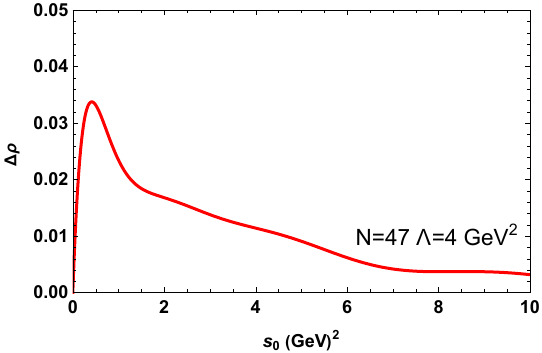}
\includegraphics[totalheight=5cm,width=7cm]{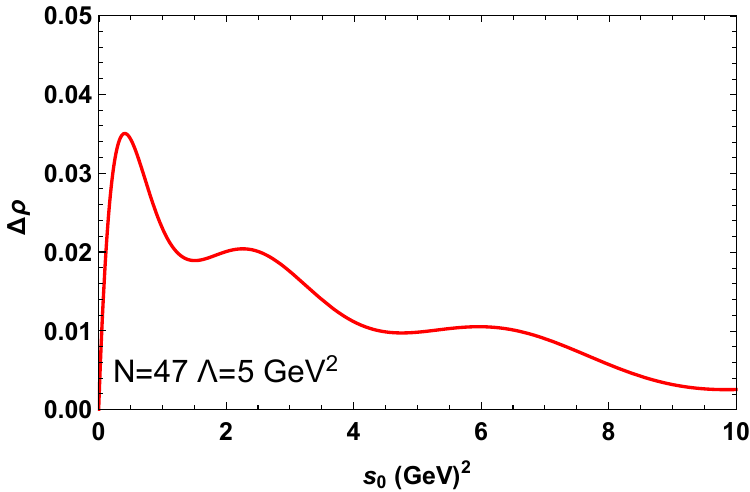}
\includegraphics[totalheight=5cm,width=7cm]{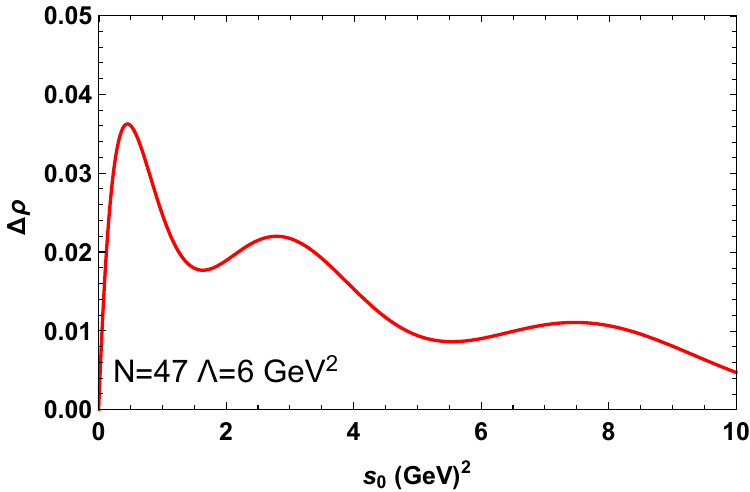}
\includegraphics[totalheight=5cm,width=7cm]{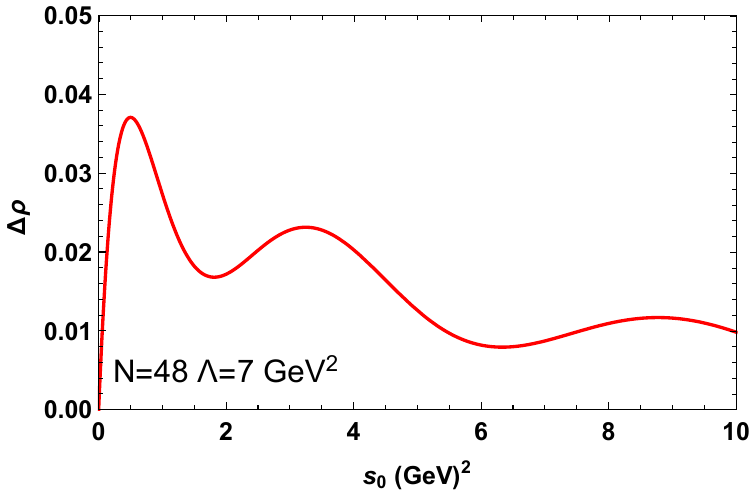}
\includegraphics[totalheight=5cm,width=7cm]{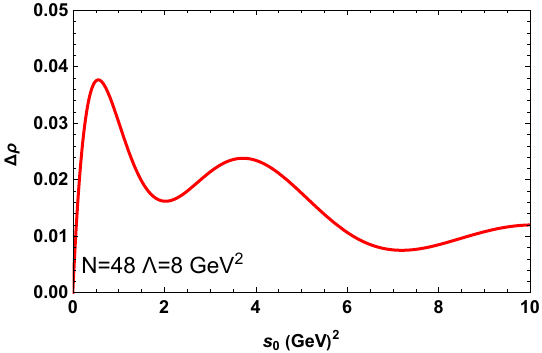}
\end{center}
\caption{Solutions of $\Delta \rho(s)$ for characteristic scales $\Lambda=1-8 \ \text{GeV}^2$ with expansion up to $N=10, 20, 30, \text{and} \ 47$ generalized Laguerre polynomials $L_n^{1}(y)$.}
\label{fig:rholambda}
\end{figure}

The spectral density $\rho(s)$ can be measured experimentally, and in principle, it should be positive or at least it should not change sign. We seek solutions to Eq. (\ref{sumrule3}) within the range $\Lambda=1-8 \ \text{GeV}^2$ for $\kappa=1$ and present the subtracted spectral densities $\Delta \rho(s)$ in Fig. \ref{fig:rholambda}. It can be seen from the figure that the positivity of the spectral density mentioned before is satisfied. The spectral densities remain positive for all $\Lambda$, which is a crucial physical requirement. Any unphysical oscillations or negative values would indicate numerical instability or inconsistencies in the reconstruction. The generalized Laguerre polynomial expansion is used to ensure a smooth reconstruction of $\Delta \rho(s)$, avoiding artificial oscillations. This suggests that the inverse QCD sum rule approach is capturing the essential features of the rho meson spectral function reliably. The other parameter is the degree $N$ which appears in Laguerre polynomial expansion. The choice of \( N \) is guided by the following principles:
\begin{enumerate}
    \item Convergence of the expansion:  
    For each \(\Lambda\), we increment \( N \) until the subtracted spectral density \(\Delta\rho(s)\) stabilizes (i.e., further increases in \( N \) do not alter the shape or peak positions of \(\Delta\rho(s)\) within numerical uncertainties). This ensures that the expansion captures the essential physics without overfitting.

    \item Positivity and smoothness:  
    Physical spectral densities must be non-negative and smooth. We discard solutions where:
    \begin{itemize}
        \item \(\Delta\rho(s)\) exhibits unphysical oscillations (indicative of overfitting),
        \item The positivity constraint \(\Delta\rho(s) \geq 0\) is violated for \(s > 0\).
    \end{itemize}
For example, in Fig. \ref{fig:rholambda} (\(\kappa = 1\)), \( N = 30 \) is chosen for \(\Lambda = 4 \ \text{GeV}^2\) because it achieves convergence while maintaining positivity.

\item Stability across \(\Lambda\):  
    The optimal \( N \) may vary with \(\Lambda\). At low \(\Lambda\) (e.g., \(1 \, \text{GeV}^2\)), fewer terms (\( N \sim 10 \)) suffice due to the dominance of non-perturbative effects. At higher \(\Lambda\) (e.g., \(8 \, \text{GeV}^2\)), more terms (\( N \sim 40 \)) are needed to resolve the perturbative continuum. 
\end{enumerate}

The spectral density graphs provide information on the mass of the rho meson ground state in nuclear matter. A prominent peak is observed around $s \simeq 0.5 \ \text{GeV}^2$, which is within the expected range for the rho meson ground state mass in nuclear matter. This peak suggests a mass shift due to medium effects, aligning with previous studies, which predict a reduction in rho meson mass in nuclear matter. It should also be mentioned that the peaks appearing in the spectral density should not necessarily represent physical states but may present information about them. The spectral density $\rho(s)$ encodes both physical states and non-resonant continuum contributions. The rho meson mass should appear as a peak in the  spectral density. It should be also mentioned that $ \text{peaks} \neq \text{direct physical states}$, i.e., not every peak in the spectral density automatically corresponds to a stable particle resonance. Some peaks may arise from (i) threshold effects, (ii) non-resonant quark-antiquark interactions and  (iii) numerical artifacts of the reconstruction method. However, the peaks carry physical information since  non-resonant peaks encode valuable QCD dynamics: (i) strength of chiral symmetry breaking ($ \left\langle \bar{q}q \right\rangle$ effects), (ii) medium-induced spectral distortions, and (iii) continuum suppression scales. Therefore the spectral density contains both resonant and non-resonant contributions, the ground state $\rho$ meson is unambiguously identified through its stable peak position and dominance in the spectral sum. Other structures reflect QCD continuum effects.

\begin{figure}[h!]
\centering
\includegraphics[scale=1]{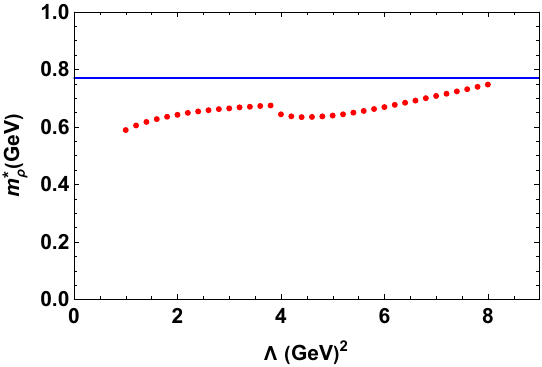}
\vspace{0.0cm} \caption{Rho meson mass $\rho_m$ dependence on $\Lambda$ for $\kappa=1$. The blue line refers to the experimental mass of the rho meson in vacuum.}
\label{masslambda}
\end{figure}

We also show the rho meson mass dependence on $\Lambda$ in nuclear matter for the range $1 \ \text{GeV}^2< \Lambda < 8 \ \text{GeV}^2$ in Fig. \ref{masslambda}. Looking at the graph, it can be seen that the mass increases in the interval $1 \ \text{GeV}^2< \Lambda < 4 \ \text{GeV}^2$, whereas the curve goes down after $\Lambda=4 \ \text{GeV}^2$ and starts to ascend monotonically as $\Lambda >4 \ \text{GeV}^2$. Furthermore, the curve in the interval $1 \ \text{GeV}^2< \Lambda < 4 \ \text{GeV}^2$ is concave, while it is convex in the interval $4 \ \text{GeV}^2< \Lambda < 8 \ \text{GeV}^2$. 

Before obtaining rho meson mass in nuclear media, we want to calculate rho meson mass in the vacuum. The rho meson mass in vacuum was calculated withtin the same framework in Refs. \cite{Li:2020ejs,Li:2021gsx,Li:2024fko} as $m_{\rho} \simeq 0.77 \ \text{GeV}$ which coincides with the mass in vacuum of the data  $m_{\rho}= 775.26 \pm 0.23 \ \text{MeV}$ \cite{ParticleDataGroup:2024cfk}. 
Using $\Lambda$ interval $2 \ \text{GeV}^2 < \Lambda < 3.8 \ \text{GeV}^2$ of the present paper, we obtain rho meson mass in vacuum as $m_{\rho}= (0.78 \pm 0.05) \ \text{GeV}$ with $\kappa=2.5$. Setting $\kappa=1$ in the same conditions yields rho meson mass as $m_{\rho} = (0.68 \pm 0.04) \  \text{GeV}$. The results are tabulated in Table \ref{comparemass}. It is clear that, turning off the medium effects and taking inputs without the medium effects in the same formalism gives higher mass values than the mass value in dense environment. 

\begin{table}[H]
\caption{Rho meson masses in vacuum and dense medium within the same formalism. The results are in unit of GeV. }\label{comparemass}
\centering
\begin{tabular}{l|c|c}
\toprule
$\kappa$ & $m_{\rho}$ (Vacuum) & $m_{\rho}^*$ (Dense medium) \\
\toprule
1& $0.68 \pm 0.04 $ & $0.66 \pm 0.03$  \\
2.5& $0.78 \pm 0.05$ & $0.70 \pm 0.04$  \\
\toprule
\end{tabular}
\end{table}

Considering the rho mass in vacuum as an asymptote in Fig. \ref{masslambda}, we evaluate the rho meson mass in nuclear matter with $\kappa=1$ between $\Lambda=2 \ \text{GeV}^2$ and $\Lambda=3.8 \ \text{GeV}^2$ as:

\begin{equation}
m_{\rho}^\ast=0.66 \pm 0.03 \ \text{GeV},
\end{equation}
where we have taken into account the uncertainties in the input parameters. The ratio of the change in the rho meson mass $\Delta m_{\rho}$ to the actual (vacuum) rho meson mass $m_{\rho}$ in nuclear matter is a measure of how significantly the mass is modified due to the medium effects. This ratio is typically expressed as:

\begin{equation}
\frac{\Delta m_{\rho}}{m_{\rho}}=\frac{m_{\rho}^\ast-m_{\rho}}{m_{\rho}} \simeq 0.1-0.2, \label{ratio}
\end{equation}
where in our case, this ratio turns out to be 0.14.

One critical assumption in QCDSR calculations is the factorization assumption, which simplifies the treatment of higher-dimensional condensates by expressing them as products of lower-dimensional condensates. This simplification is based on the vacuum saturation hypothesis, which assumes that the vacuum state dominates the contributions to these condensates. While this assumption makes calculations tractable, its validity in nuclear matter is not guaranteed, and deviations from factorization can significantly impact predictions of in-medium hadronic properties. Furthermore, in nuclear matter, the presence of nucleons and their interactions can lead to significant deviations from vacuum saturation. The medium effects, such as partial restoration of chiral symmetry, modify the quark and gluon condensates, potentially breaking the factorization assumption. This discussion focuses on the impacts of the factorization assumption on the rho meson mass in nuclear matter, highlighting its theoretical and phenomenological implications.

\begin{figure}[h!]
\begin{center}
\includegraphics[totalheight=5cm,width=7cm]{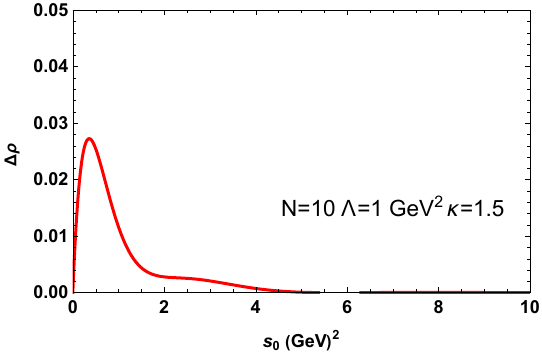}
\includegraphics[totalheight=5cm,width=7cm]{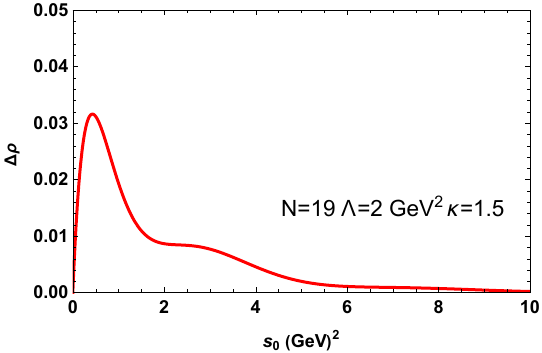}
\includegraphics[totalheight=5cm,width=7cm]{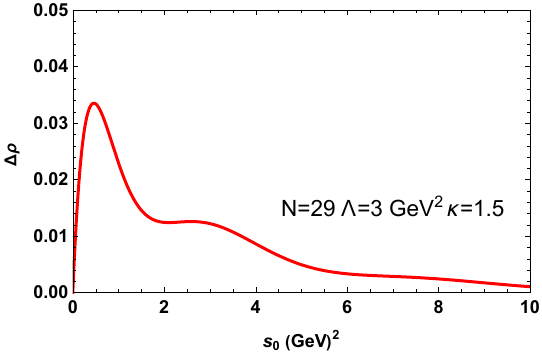}
\includegraphics[totalheight=5cm,width=7cm]{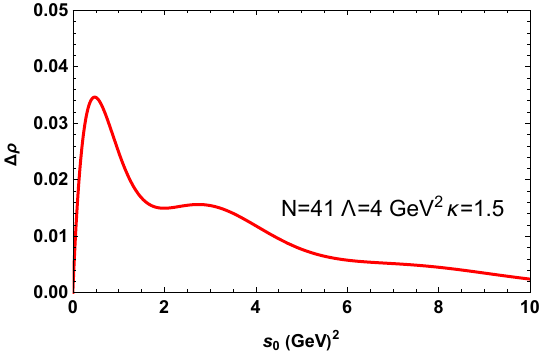}
\includegraphics[totalheight=5cm,width=7cm]{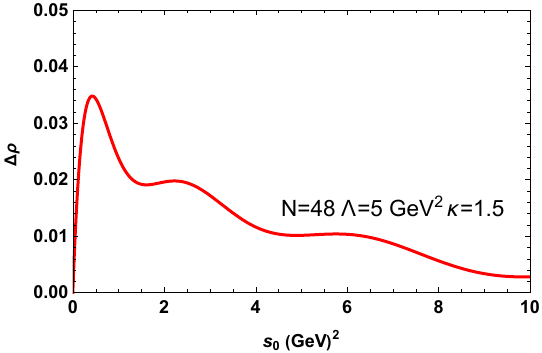}
\includegraphics[totalheight=5cm,width=7cm]{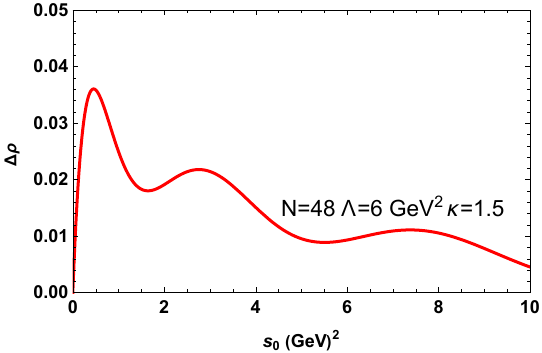}
\includegraphics[totalheight=5cm,width=7cm]{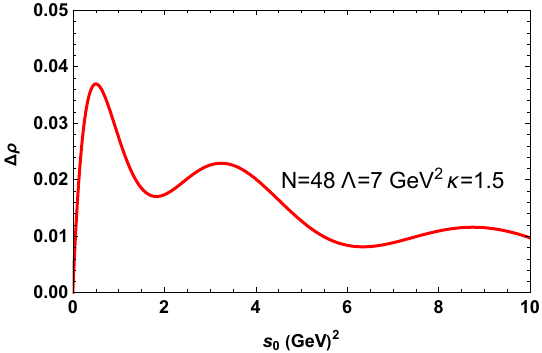}
\includegraphics[totalheight=5cm,width=7cm]{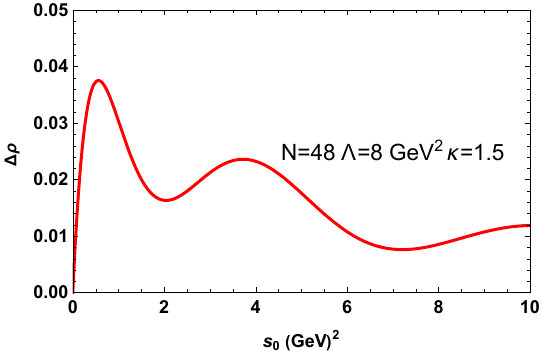}
\end{center}
\caption{Solutions of $\Delta \rho(s)$ for characteristic scales $\Lambda=1-8 \ \text{GeV}^2$ with expansion up to $N=10, 19, 41, \text{and} \ 48$ generalized Laguerre polynomials $L_n^{1}(y)$ for $\kappa=1.5$.}
\label{fig:rholambda1}
\end{figure}

\begin{figure}[h!]
\begin{center}
\includegraphics[totalheight=5cm,width=7cm]{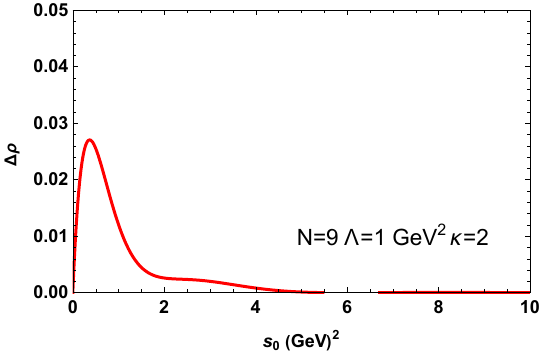}
\includegraphics[totalheight=5cm,width=7cm]{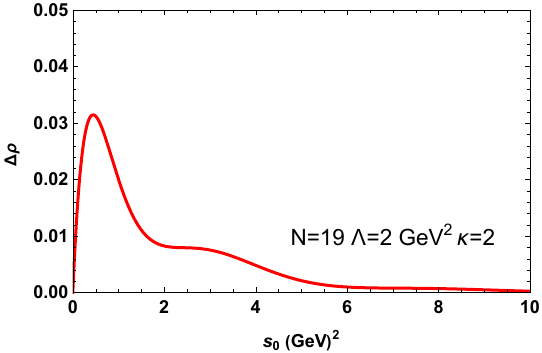}
\includegraphics[totalheight=5cm,width=7cm]{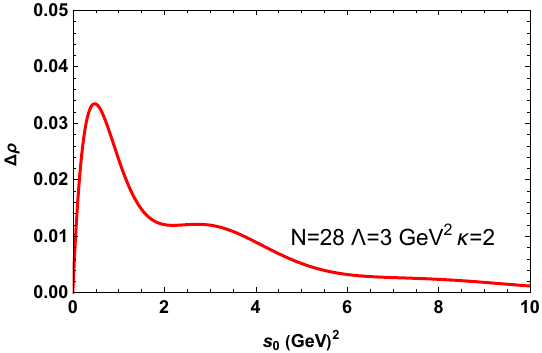}
\includegraphics[totalheight=5cm,width=7cm]{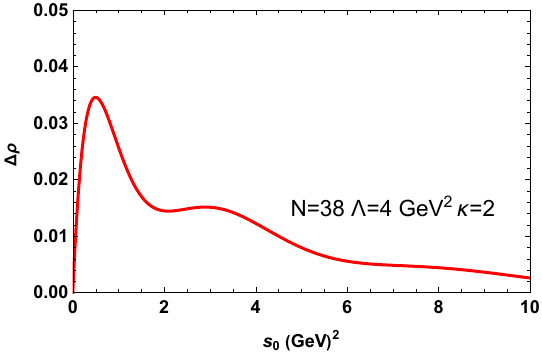}
\includegraphics[totalheight=5cm,width=7cm]{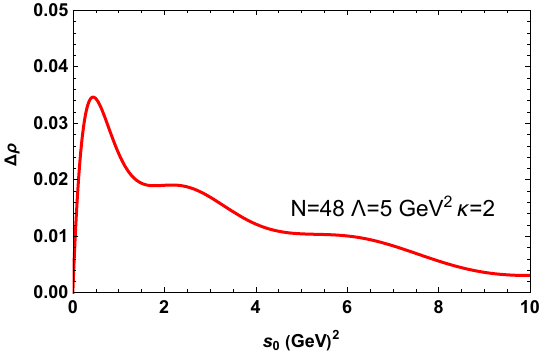}
\includegraphics[totalheight=5cm,width=7cm]{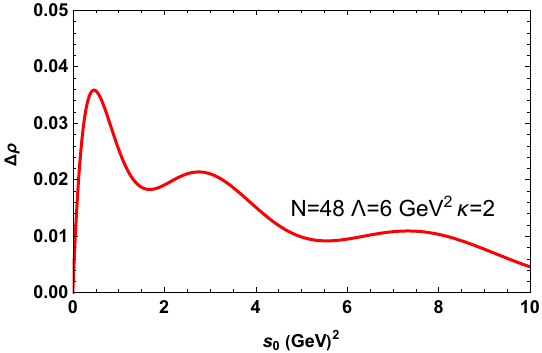}
\includegraphics[totalheight=5cm,width=7cm]{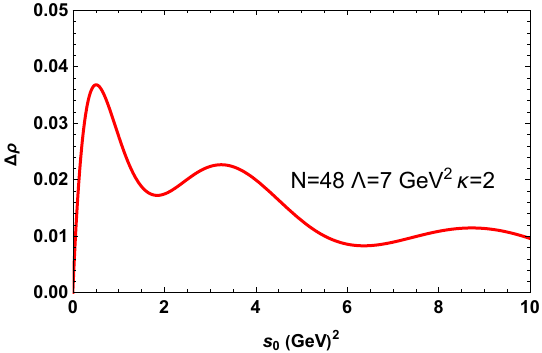}
\includegraphics[totalheight=5cm,width=7cm]{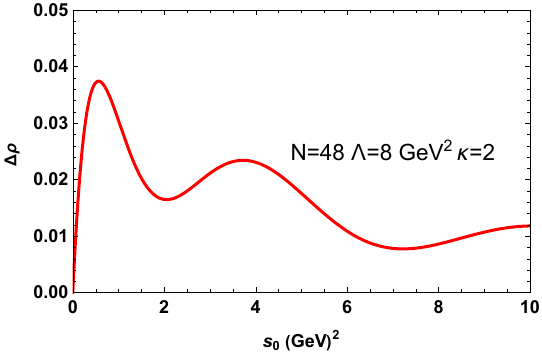}
\end{center}
\caption{Solutions of $\Delta \rho(s)$ for characteristic scales $\Lambda=1-8 \ \text{GeV}^2$ with expansion up to $N=9, 19, 28, 38, \text{and} \ 48$ generalized Laguerre polynomials $L_n^{1}(y)$ for $\kappa=2$.}
\label{fig:rholambda2}
\end{figure}

\begin{figure}[h!]
\begin{center}
\includegraphics[totalheight=5cm,width=7cm]{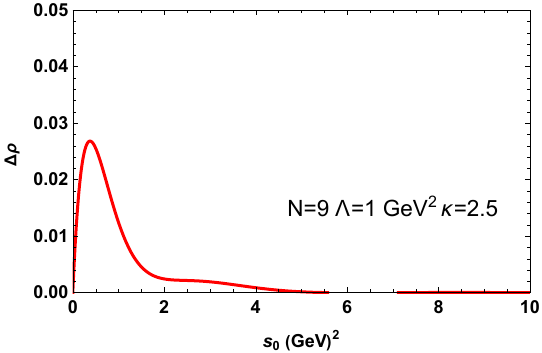}
\includegraphics[totalheight=5cm,width=7cm]{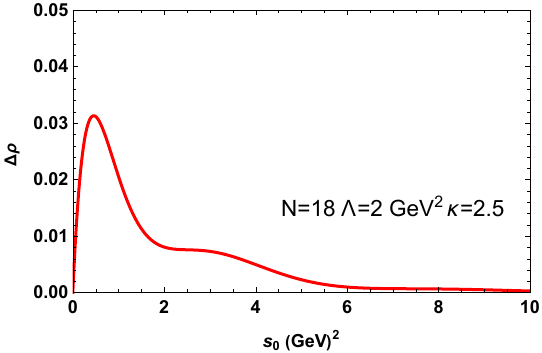}
\includegraphics[totalheight=5cm,width=7cm]{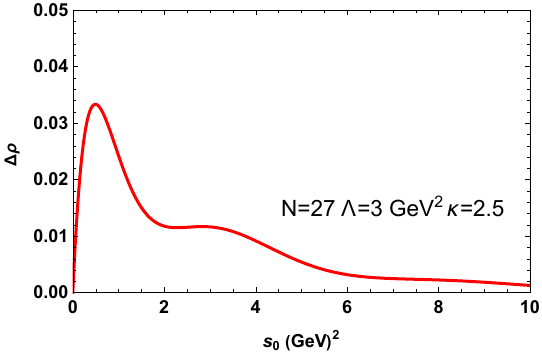}
\includegraphics[totalheight=5cm,width=7cm]{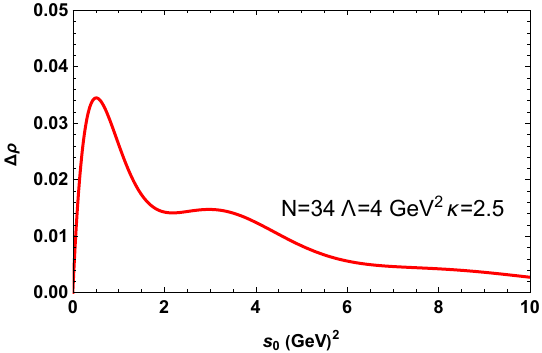}
\includegraphics[totalheight=5cm,width=7cm]{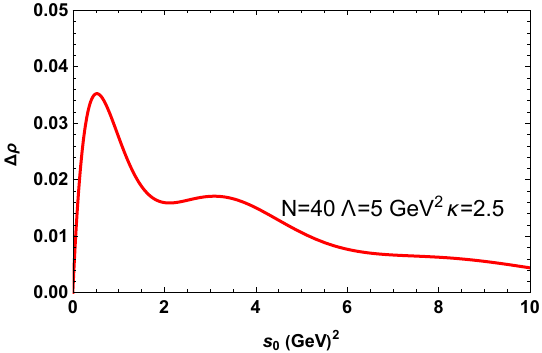}
\includegraphics[totalheight=5cm,width=7cm]{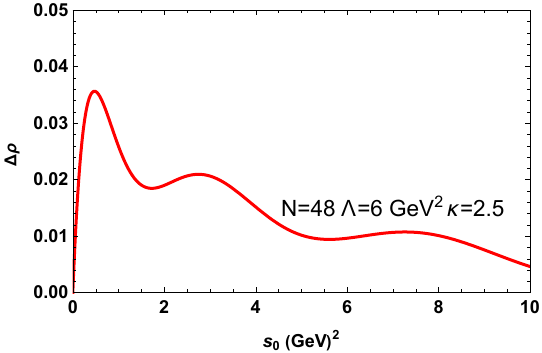}
\includegraphics[totalheight=5cm,width=7cm]{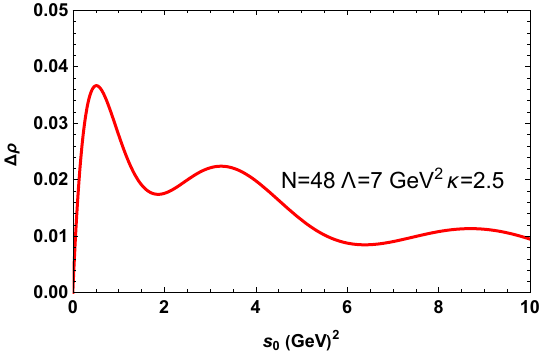}
\includegraphics[totalheight=5cm,width=7cm]{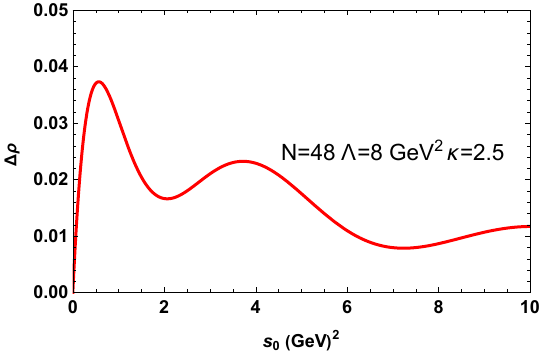}
\end{center}
\caption{Solutions of $\Delta \rho(s)$ for characteristic scales $\Lambda=1-8 \ \text{GeV}^2$ with expansion up to $N=9, 18, 27, 34, 40, \text{and} \ 48$ generalized Laguerre polynomials $L_n^{1}(y)$ for $\kappa=2.5$.}
\label{fig:rholambda3}
\end{figure}

\begin{figure}[h!]
\begin{center}
\includegraphics[totalheight=5cm,width=7cm]{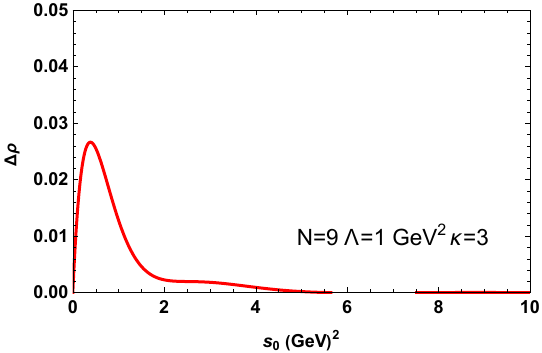}
\includegraphics[totalheight=5cm,width=7cm]{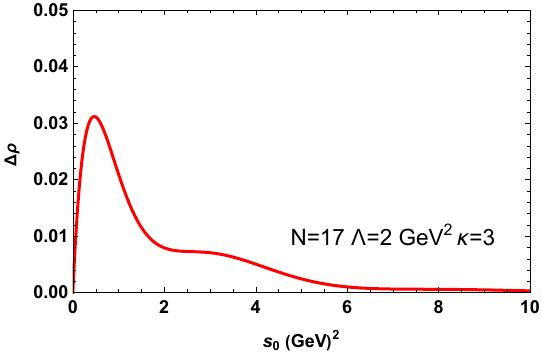}
\includegraphics[totalheight=5cm,width=7cm]{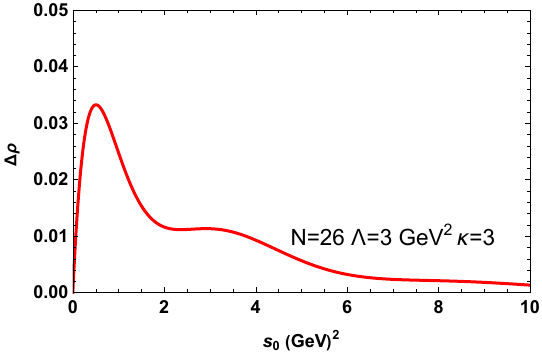}
\includegraphics[totalheight=5cm,width=7cm]{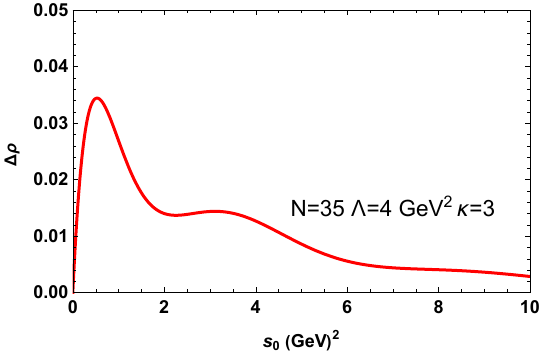}
\includegraphics[totalheight=5cm,width=7cm]{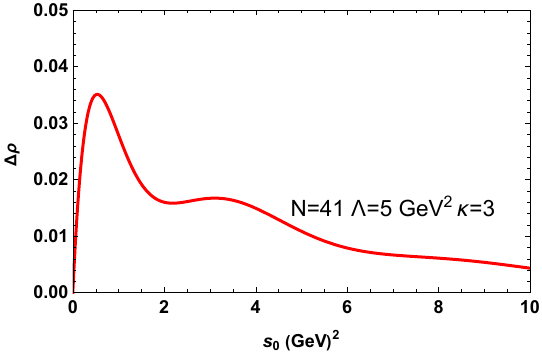}
\includegraphics[totalheight=5cm,width=7cm]{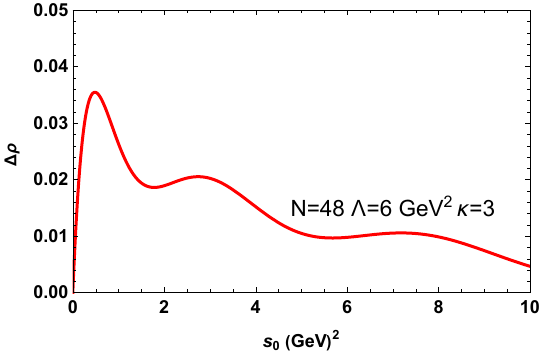}
\includegraphics[totalheight=5cm,width=7cm]{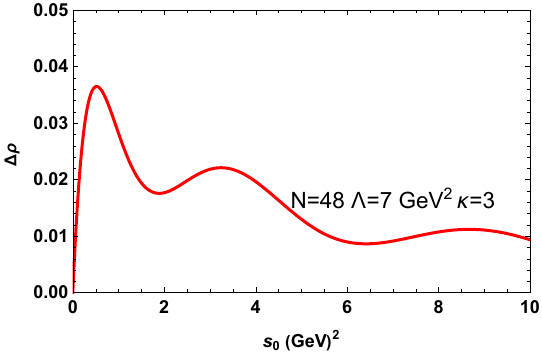}
\includegraphics[totalheight=5cm,width=7cm]{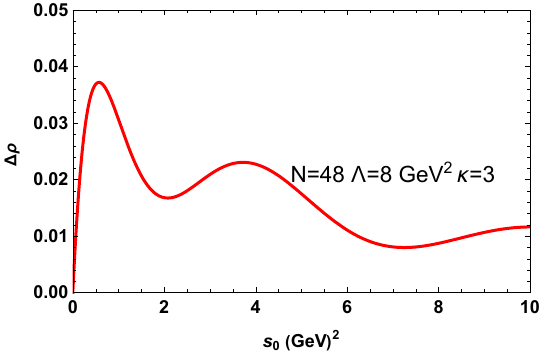}
\end{center}
\caption{Solutions of $\Delta \rho(s)$ for characteristic scales $\Lambda=1-8 \ \text{GeV}^2$ with expansion up to $N=9, 17, 26, 35, 41, \text{and} \ 48$ generalized Laguerre polynomials $L_n^{1}(y)$ for $\kappa=3$.}
\label{fig:rholambda4}
\end{figure}

\begin{figure}[h!]
\begin{center}
\includegraphics[totalheight=5cm,width=7cm]{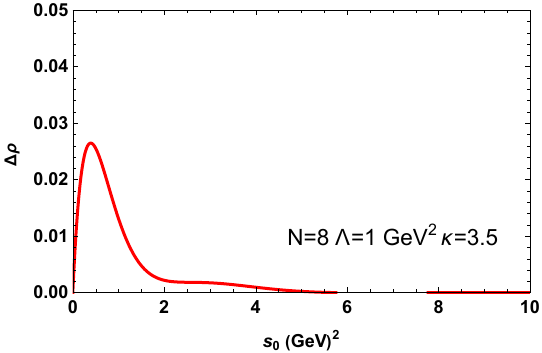}
\includegraphics[totalheight=5cm,width=7cm]{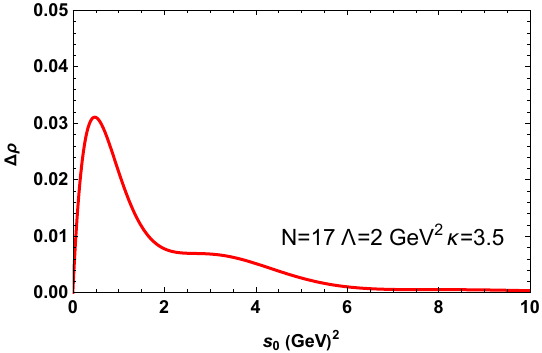}
\includegraphics[totalheight=5cm,width=7cm]{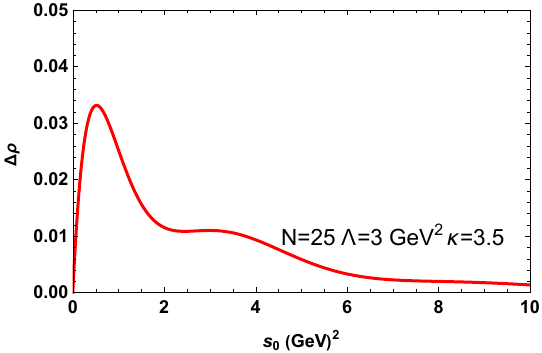}
\includegraphics[totalheight=5cm,width=7cm]{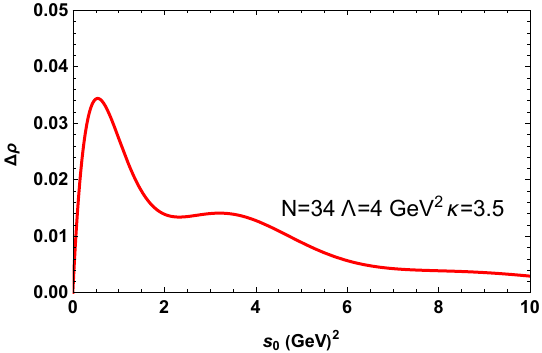}
\includegraphics[totalheight=5cm,width=7cm]{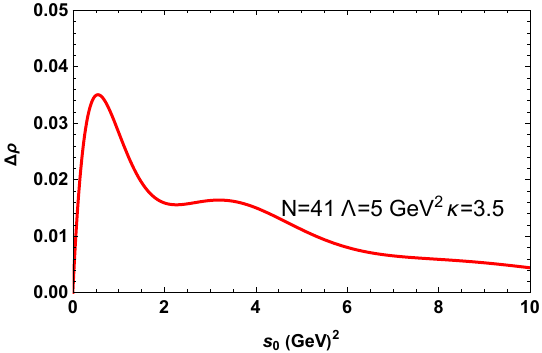}
\includegraphics[totalheight=5cm,width=7cm]{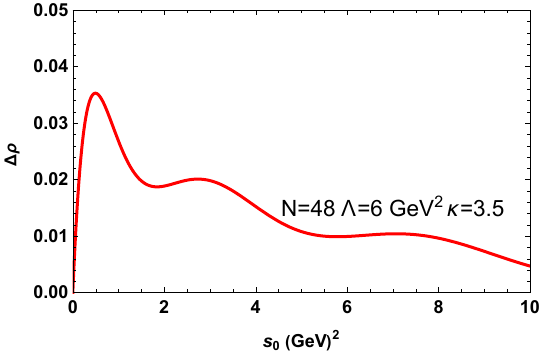}
\includegraphics[totalheight=5cm,width=7cm]{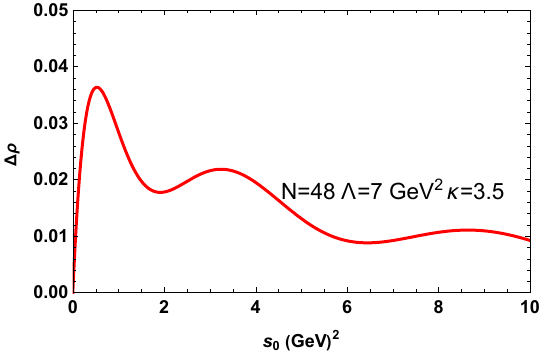}
\includegraphics[totalheight=5cm,width=7cm]{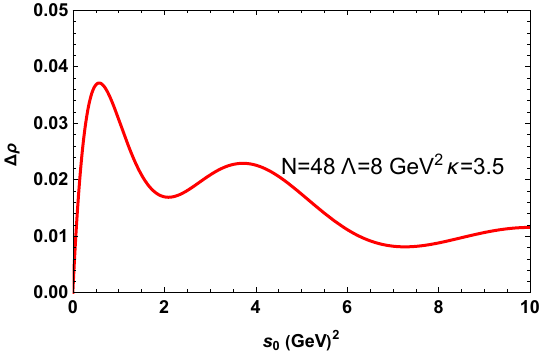}
\end{center}
\caption{Solutions of $\Delta \rho(s)$ for characteristic scales $\Lambda=1-8 \ \text{GeV}^2$ with expansion up to $N=8, 17, 25, 34, 41, \text{and} \ 48$ generalized Laguerre polynomials $L_n^{1}(y)$ for $\kappa=3.5$.}
\label{fig:rholambda5}
\end{figure}

We also investigate the role of the factorization assumption, which is commonly employed in QCDSR but lacks rigorous justification in a nuclear medium. In order to see the factorization effects, we make a comprehensive analysis. In Figs. \ref{fig:rholambda1}, \ref{fig:rholambda2}, \ref{fig:rholambda3}, \ref{fig:rholambda4}, and \ref{fig:rholambda5}, subtracted spectral densities for $\kappa=1.5$, $\kappa=2$, $\kappa=2.5$, $\kappa=3$, and $\kappa=3.5$ are presented, respectively. The positivity of the spectral density is lost for $\Lambda=1 \ \text{GeV}^2$ across all values of $\kappa$, indicating that at very low scales, the inverse QCDSR method struggles to maintain a physically meaningful spectral function. However, while the subtracted spectral density \(\Delta\rho(s)\) for \(\Lambda = 1\,\text{GeV}^2\) exhibits regions of negativity (Figs. \ref{fig:rholambda1}, \ref{fig:rholambda2}, \ref{fig:rholambda3}, \ref{fig:rholambda4}, \ref{fig:rholambda5}), this does not indicate a physical inconsistency. The original spectral density \(\rho(s)\), reconstructed via Eq. \ref{subtracted}, satisfies \(\rho(s) \geq 0\) for all \(s > 0\) and \(\kappa\) values. The negativity in \(\Delta\rho(s)\) arises from the subtraction of the perturbative baseline, which is larger than \(\rho(s)\) in certain kinematic regions. This behavior is expected and does not affect the physical interpretation of the results. For $\Lambda > 1 \ \text{GeV}^2$, the spectral densities are positive and stable, meaning the factorization assumption still holds reasonably well in nuclear matter at higher scales. A prominent peak is observed in the spectral densities around $s \simeq 0.5 \ \text{GeV}^2$, which is consistent with the expected mass of the rho meson in nuclear matter. As mentioned before, the factorization assumption simplifies the treatment of higher-dimensional condensates by expressing them as products of lower-dimensional condensates. While this assumption holds reasonably well at higher scales $\Lambda > 1 \ \text{GeV}^2$, the breakdown at $\Lambda = 1 \ \text{GeV}^2$ suggests that medium effects (such as partial restoration of chiral symmetry) may significantly modify the quark and gluon condensates, potentially breaking the factorization assumption at low scales. The positivity of the spectral density $\rho(s)$, observed for all $\Lambda$ and $\kappa$ values, reflects the physical requirement of unitarity. This condition is maintained regardless of whether the factorization assumption holds ($\kappa=1$) or is violated ($\kappa>1$). The factorization-breaking effects, parameterized by $\kappa>1$, manifest in the detailed structure of the OPE coefficients rather than in the overall positivity of the spectral function. 

In this formalism, the decay constants can be obtained as follows. In Ref. \cite{CP-PACS:2001ncr}, it was highlighted that the area under the resonance peak of the spectral density $\rho(s)$ is equivalent to the square of the decay constant. In present formulation, the resonance peak is approximately represented by the subtracted spectral density where continuum contribution is largely removed. Therefore it is possible to write
\begin{equation}
f_{\rho}^2 \approx \int_0^\infty \Delta \rho (s,\Lambda) ds. \label{decayconstant}
\end{equation}

We list the masses and decay constants of the rho meson for the values of $\kappa$ in Table \ref{massdecay}. As clear from the table, the mass reduction ranges from 6\% to 14\%. For $\kappa=1,1.5,2$, and $\kappa=2.5$ values, the mass shift ratio is consistent with the range of mass shifts (10-20\%) reported in the literature for the rho meson in nuclear matter. The ratio decreases as $\kappa$ increases, indicating that the mass reduction becomes smaller for higher values of $\kappa$. Our results show that deviations from factorization can significantly impact the extracted hadron properties, highlighting the need for improved nuclear medium corrections. 

\begin{table*}[h!]
\caption{Masses and decay constants of the rho meson in nuclear matter with respect to $\kappa$. The results are presented in units of GeV.}\label{massdecay}
\begin{tabular}{l|c|c|c}
\toprule
$\kappa$ & $m_\rho^*$ & $f^*_\rho$ &$\Delta m_\rho/m_\rho$ \\
\toprule
1 & $0.66 \pm 0.03$ & $0.102 \pm 0.08$ & 0.14 \\
1.5 & $0.67 \pm 0.03$ & $0.102 \pm 0.08$ & 0.13 \\
2 & $0.69 \pm 0.04$ & $0.103 \pm 0.09$ & 0.10 \\
2.5 & $0.70 \pm 0.04$ & $0.103 \pm 0.09$ & 0.09 \\
3 & $0.71 \pm 0.04$ & $0.104 \pm 0.09$ & 0.07 \\
3.5 & $0.72 \pm 0.05$ & $0.105 \pm 0.09$ & 0.06 \\
\toprule
\end{tabular}
\end{table*}

\section{Final Remarks}\label{final}

The study of rho mesons in nuclear matter is a significant area of research in nuclear physics, providing insights into the behavior of mesons under extreme conditions. As vector mesons, rho mesons exhibit notable modifications in their properties when immersed in dense nuclear environments. These modifications include changes in mass, width, and spectral functions, which are crucial for understanding the interactions within nuclear matter.

In traditional QCDSR, the spectral density is usually parameterized as a pole+perturbative continuum. This method relies on an arbitrary continuum threshold $s_0$, which can introduce model dependence. The inverse QCDSR method, on the other hand, reconstructs the spectral function directly from the OPE, avoiding assumptions about the pole + continuum structure. In the inverse QCDSR method, the spectral density is reconstructed directly from the OPE, without assuming a specific form (e.g., a pole + continuum model).

In this work, we have revisited the properties of the rho meson in nuclear matter using the inverse QCD sum rules method. Our results show a significant reduction in the rho meson mass, consistent with the 10-20\% reduction predicted by previous studies such as those in \cite{Hatsuda:1991ez,Asakawa:1993pq,Hatsuda:1995dy}. The broadening of the spectral function, observed in our analysis, aligns with findings in Refs. \cite{Rapp:1997fs,Klingl:1997kf}, which emphasized the role of medium effects and baryon resonances in modifying the spectral density. This approach, free from the conventional assumptions about spectral density parameterization, offers an independent confirmation of the well-established mass shift and broadening effects of the rho meson in dense environments. The inverse QCDSR method has demonstrated its strength in reconstructing the spectral function directly from the OPE, revealing a significant reduction in the rho meson mass, consistent with previous theoretical predictions.

Furthermore, our findings emphasize the importance of medium effects, including the partial restoration of chiral symmetry and the influence of higher-order condensates. The sensitivity of the results to the factorization assumption highlights the necessity of refining nuclear medium corrections for a more precise determination of hadronic properties in dense environments. This study not only corroborates the validity of the inverse QCDSR method but also paves the way for its broader application in exploring in-medium modifications of other hadrons.

This study reaffirms the utility of the inverse QCD sum rule method as a powerful tool for investigating hadron modifications in nuclear matter. The inverse QCDSR formalism is a powerful and systematic method for extracting hadronic properties from QCD. It avoids assumptions about the spectral density and directly utilizes the underlying theory through dispersion relations and the OPE. This makes it a valuable tool for exploring non-perturbative aspects of QCD and understanding the structure of hadrons. Future work could extend this formalism to investigate excited states and exotic hadronic configurations in nuclear matter, providing deeper insights into the strong interaction under extreme conditions. Our results underline the significance of nonperturbative QCD techniques in unraveling the complex structure of hadronic matter and contribute to the ongoing efforts to bridge theoretical predictions with experimental observations in heavy-ion collisions and nuclear astrophysics.

\bibliography{rhomeson-revised}

\end{document}